\documentclass[mathleft
]{an}
\usepackage{graphicx}
\usepackage{times}
\begin{document}

% The following seven commands are intended for editorial usage and should be ignored by
% the author(s).
\Pagespan{1}{6}% Document's page range. 
% If second parameter is left empty, the last page is computed automatically.
\Yearpublication{}%
\Yearsubmission{2011}%
\Month{}%   
\Volume{}%  
\Issue{}% 
% \DOI{This.is/not.aDOI}% 

%\topmargin+0.5cm

\title{Angular momentum transport by thermal emission in black hole accretion disks}

\author{Jarrett L. Johnson\thanks{
  \email{jjohnson@mpe.mpg.de}\newline}
%Example 
%for footnote, note the usage of the \texttt{fnmsep}
%command as separator between institute number and footnote mark}
}
\titlerunning{Angular momentum transport by thermal emission}
\authorrunning{J.~L. Johnson}
\institute{Max-Planck-Institut f{\"u}r extraterrestrische Physik, \\ 
Giessenbachstra\ss{}e, 85748 Garching, Germany \\ \\
Los Alamos National Laboratory \\
Los Alamos, NM 87545, USA}

\received{}
\accepted{}
\publonline{}


\abstract{%
  We calculate the amount of angular momentum that thermal photons carry out of a viscous black hole accretion disk, due to the strong Doppler shift imparted to them 
by the high orbital velocity of the radiating disk material.  While thermal emission can not drive accretion on its own, we show that along with disk heating it does 
nonetheless result in a loss of specific angular momentum, thereby contributing to an otherwise viscosity-driven accretion flow.  In particular, 
we show that the fraction of the angular momentum that is lost to thermal emission at a radius $r$ in a standard, multi-color disk is $\sim$ 0.4$r_{\rm s}$/$r$, 
where $r_{\rm s}$ is the Schwarzschild radius of the black hole.  We briefly highlight the key similarties between this effect and the closely related Poynting-Robertson effect.
}

\maketitle

\section{Introduction}
The process by which matter is accreted onto black holes is a topic of long-standing 
and continued interest (for reviews see e.g. Pringle 1981; Armitage 2004).  An issue of central importance
is how the angular momentum of the accreting material is lost within the accretion disk
that forms around a black hole, as it is the loss of angular momentum which enables material to pass through
the disk and to fall into the black hole.  Numerous processes are likely to contribute to the transport of angular momentum, 
including particle viscosity (see e.g. Rossi \& Olbert 1970), kinematic viscosity generated by turbulence (e.g. Shakura \& Sunyaev 1973), 
magnetohydrodynamic instability (e.g. Balbus \& Hawley 1998), and radiative viscosity in shearing accretion flows (Loeb \& Laor 1992).  
%In addition, radiation that escapes from an accretion disk will carry away angular momentum and 
%contribute to driving the accretion flow (e.g. Page \& Thorne 1974).

As expected from simple theoretical considerations (see e.g. Pringle 1981), one of the common hallmarks of accretion disks 
is a strong thermal (multi-color) component of the emitted spectrum, as has been observed from the disks around stellar 
mass (e.g. Mitsuda et al. 1984; Makishima et al. 2000), intermediate mass (e.g. Farrell et al. 2009), and supermassive 
black holes (e.g. Yuan et al. 2010).  The prodigious energy that is radiated away is accompanied by a correspondingly large amount of momentum 
and, in addition, a significant amount of angular momentum (e.g. Page \& Thorne 1974).  Importantly, however, an accretion flow can not 
be driven solely by the isotropic emission of radiation.  An additional ingredient that is required is a source of heating to balance the loss
of energy to the emitted radiation, as we discuss in the present work.   
 
In the following Sections, we carry out the first explicit calculation of the amount of angular momentum that is \linebreak
transported out of a standard multi-color accretion disk due to viscous heating and thermal emission, thereby helping to facilitate the accretion of material onto the central black hole.  
In the following Section we calculate the timescale on which angular momentum is transported out of the disk by thermal emission, and in Section 3 
we calculate the fraction of the angular momentum lost in an accretion disk that is carried away by thermal photons.  Finally, we 
summarize our findings and give our concluding remarks in Section 4.

\begin{figure}
  \centering
  \includegraphics[width=2.68in]{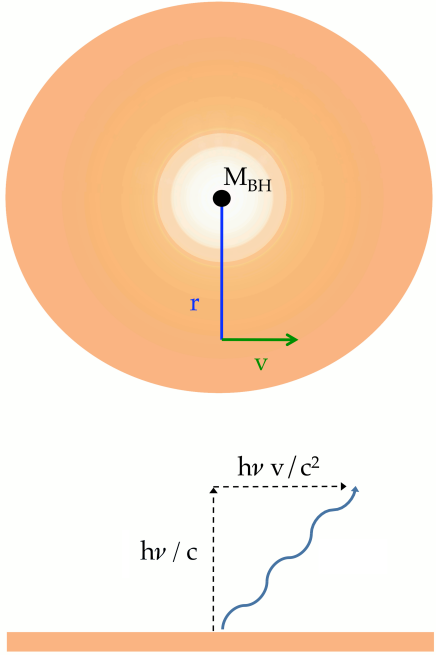}
  \caption
  {Schematic depiction of the viscous, multi-color accretion disk under consideration, shown face-on ({\it top}) and edge-on ({\it bottom}).  
At radius $r$ the material orbits the central black hole at velocity $v$, as defined by equation (2).  In the frame of the disk, 
a photon with energy $h\nu$ is emitted from radius $r$ in the direction perpendicular to the disk. In the lab frame, the photon carries a momentum $h\nu$/$c$ in the 
direction perpendicular to the disk, and a momentum $h\nu$$v$/c$^2$ in the direction of motion of the disk.  Thus, such a photon will carry away an amount $l$ = $h\nu$$v$$r$/$c^2$ 
of the angular momentum in the disk, as given by equation (3).  While here we show only a single photon, if the emission is isotropic, the average angular momentum carried by a 
photon is the same as given by equation (3).  In the comoving frame of the disk, the loss of specific angular momentum is attributable to the gain in mass/energy that is 
imparted due to viscous heating, which balances the cooling due to isotropic thermal emission (see also e.g. Robertson 1937).}
\end{figure}

\section{Angular momentum transport by thermal photons}
Here we describe the process of angular momentum transport by viscous heating and thermal emission, and 
we derive an expression for the timescale on which this process acts in an accretion disk.  
To begin, consider a standard, viscous multi-color accretion disk around a black hole of mass $M_{\rm BH}$, as shown in Figure 1.  
The temperature of the disk at a given distance 
from the central black hole is set by the balance between viscous heating and radiative cooling; as a function of the distance $r$ 
from the black hole, the temperture profile 
is thus found to be well approximated by (e.g. Pringle 1981)

\begin{equation}
T = \left(\frac{3 G M_{\rm BH} \dot{M_{\rm BH}}}{8 \pi \sigma r^3}\right)^{\frac{1}{4}}            {\mbox ,}
\end{equation}
where $\sigma$ is the Stefan-Boltzmann constant, $G$ is Newton's constant, and 
$\dot{M_{\rm BH}}$ is the steady-state accretion rate of the black hole (and also of material passing through the disk).
Also, take it that the disk is approximately Keplerian, i.e. that the velocity at which material orbits the black hole is given by (e.g. Verbunt 1982)

\begin{equation}
v = \left(\frac{G M_{\rm BH}}{r} \right)^{\frac{1}{2}} {\mbox .}
\end{equation}

Now, consider the angular momentum $l$ that is carried away by a photon with energy $h \nu$ that is emitted from the accretion disk at radius $r$, as shown in Fig. 1:

\begin{equation}
l = \frac{h \nu v r}{c^2} {\mbox ,}
\end{equation}
where $c$ is the speed of light.  For the case of a photon emitted in the direction perpendicular to the disk (in the frame of the disk), as shown in Fig. 1, a component of the momentum of the photon is in the direction 
of motion of the disk, in the lab frame.  This is due essentially to the Doppler effect:  photons emitted in the 
direction of the motion of the disk will have higher energy, while photons emitted in the opposite direction will have 
lower energy (see also Robertson 1937 and references therein).  
For monochromatic radiation emitted isotropically from the disk, the average angular momentum carried away by a photon is given by equation (3). In turn, as angular momentum must be conserved,
the radiating material in the disk experiences a net loss of angular momentum (see e.g. Page \& Thorne 1974).  
This effects a loss of centripetal support and an enhancement in the rate at which material falls towards the black hole.

As noted in Robertson (1937; e.g. pages 424-425), a source of heating to balance the loss of energy to radiation is essential to this effect. Without it there would be a net loss 
of mass/energy to the emitted radiation, and while angular momentum would be lost, there would hence be no loss of {\it specific} angular momentum and thus no infall (see also Hsieh \& Spiegel 1976).
In the accretion disk under consideration, the heating is provided by viscosity, as described in e.g. Pringle (1981).  This leads to the temperature of a fluid element in the disk increasing 
as it falls to smaller radii, with $T$ $\propto$ $r^{-\frac{3}{4}}$, as given by equation (1).  This increase in temperature implies that there is indeed no net loss of mass/energy from an 
infalling fluid element, but rather an increase. Thus, a fluid element must experience an overall loss of specific angular momentum due to the emission of radiation, 
which in turn has the effect of speeding its infall toward the central black hole.  For illustrative purposes, a brief comparison of the effect considered in present work with the classical case
considered by Robertson (1937) and Poynting (1903) is given in the appendix.

As the total energy radiated away per unit time per unit area is 
$\sigma$$T^4$ for a thermal black body, following equation (3) we arrive at a formula for the rate at which 
the total angular momentum $L$ of the disk is carried away by thermal radiation emitted at radius $r$:

\begin{equation}
\frac{dL}{dr dt} = \frac{2 \pi r^2 v \sigma T^4}{c^2} {\mbox .}
\end{equation}
This becomes, using equations (1) and (2),

\begin{equation}
\frac{dL}{drdt} = \frac{3 \dot{M_{\rm BH}}}{4 c^2} \left(\frac{G M_{\rm BH}}{r} \right)^{\frac{3}{2}} {\mbox .} 
\end{equation}

Now, the total angular momentum $L$ in the disk varies with radius as

\begin{equation}
\frac{dL}{dr} = 2 \pi r^2 v \Sigma {\mbox ,}
\end{equation}
where $\Sigma$($r$) is the surface density of the disk at radius $r$.  Again using equation (2), this becomes

\begin{equation}
\frac{dL}{dr} = 2 \pi G^{\frac{1}{2}}  M_{\rm BH}^{\frac{1}{2}} r^{\frac{3}{2}} \Sigma {\mbox .}
\end{equation}
Dividing equation (7) by equation (5) yields an estimate of the timescale $t_{\rm acc}$ for angular momentum to be transported out of the disk 
at radius $r$ by thermal emission:

\begin{eqnarray}
t_{\rm acc} & \sim & \frac{8 \pi c^2 r^3 \Sigma}{3 G M_{\rm BH} \dot{M_{\rm BH}}}  \nonumber \\
          &  \sim &  10^{3} {\rm yr} \left(\frac{\dot{M_{\rm BH}}}{{\rm M_{\odot} yr^{-1}}} \right)^{-1} \left(\frac{M_{\rm BH}}{10^{9} {\rm M_{\odot}}} \right)^{2} \left(\frac{r}{10 r_{\rm s}} \right)^{3} \nonumber \\
& \times &  \left(\frac{\Sigma}{{\rm 10^4 g cm^{-2}}} \right) {\mbox ,} 
\end{eqnarray}
where we have used $r_{\rm s}$ = 2 $GM_{\rm BH}$/c$^2$.

We note that, as it is radiation escaping the accretion disk that transports angular momentum, it is only from the photosphere of the disk
that angular momentum will be lost.  In effect, however, in our calculation of $t_{\rm acc}$
we have assumed that the material in the disk is mixed well enough that the specific angular momentum of material radiating at the photosphere is equal 
to the vertically averaged specific angular momentum.
To check this assumption, we estimate the timescale on which material from the equatorial region of the disk is transported 
to the photosphere by turbulent motion, as follows. 
The turbulent Mach number of the material is given by $M$ $\equiv$ $v_{\rm turb}$ / $c_{\rm s}$.  Then, we have the timescale $t_{\rm mix}$ on 
which material is mixed through the vertical extent of the disk, taken to be the scale height $H$, as

\begin{equation}
t_{\rm mix} \sim \frac{H}{M c_{\rm s}} = \frac{r}{M v} \sim \frac{t_{\rm Kep}}{M} \mbox{\ ,}
\end{equation}
where again $v$ is given by equation (2) and $t_{\rm Kep}$ = 2$\pi$$r$/$v$ is the orbital timescale.  
Even for a mildly turbulent disk (i.e. $M$ $\la$ 1), $t_{\rm Kep}$ $\la$ $t_{\rm mix}$ and angular momentum 
will be distributed throughout the vertical extent of the disk during the course of, at most, a few orbits.  
Therefore, we conclude that our assumption of a uniform specific angular momentum distribution is in general a sound one.

\section{Efficacy for driving an accretion flow}
Here we consider the degree to which angular momentum transport by thermal emission can contribute to driving an accretion flow.  
Using energetic and timescale arguments we show that while a significant amount of the angular momentum lost in an accretion disk
is lost to radiation, this process is most effective at enhancing the accretion flow at small radii, near the event horizon of the central black hole.

\subsection{The gravitational potential energy budget}
A fundamental limit on the angular momentum that can be carried away by thermal emission is set by the gravitational potential 
energy that is available to the gas in the disk as it falls inwards.  Here we calculate the maximum fraction of the angular momentum 
that is lost by a fluid element in passing through the disk that can be carried away by thermal emission.  

From the virial theorem, the maximum amount of energy $E_{\rm max}$ that can be extracted from an infalling fluid element of mass $m$ at a radius $r$ 
is half the absolute value of the gravitational potential energy: $E_{\rm max}$ = $G$$M_{\rm BH}$$m$/2$r$.  If this energy is converted entirely into
radiation that is emitted during the infall of the fluid element through the accretion disk from radius $r$+$dr$ to $r$ then, following equation (3), the 
maximum angular momentum $dl_{\rm rad}$ that can be radiated away is given by

\begin{equation}
dl_{\rm rad} = \frac{1}{2}\frac{m}{c^2}\left(\frac{GM_{\rm BH}}{r}\right)^{\frac{3}{2}}dr {\mbox .}
\end{equation}
In turn, the angular momentum $dl_{\rm req}$ that must be lost by the fluid element, orbiting at approximately 
the Keplerian velocity $v$ (equation 2), in falling the distance $dr$ is given by 

\begin{equation}
dl_{\rm req} = \frac{1}{2}m\left(\frac{GM_{\rm BH}}{r}\right)^{\frac{1}{2}} dr {\mbox .}
\end{equation}
The ratio of these two quantities gives the maximum ratio of the angular momentum lost at radius $r$ that can be carried away by radiation:

\begin{equation}
\frac{dl_{\rm rad}}{dl_{\rm req}} =  \frac{1}{2} \frac{r_{\rm s}}{r} {\mbox ,}
\end{equation}
where again $r_{\rm s}$ is the Schwarzschild radius of the black hole.  Therefore, due to the limited amount of energy that is available for 
radiation, the fraction of angular momentum that is carried away via radiation is relatively small, but not negligible.  For example, at $r$ = 10$r_{\rm s}$ roughly
5 percent of the angular momentum is radiated away, but at the innermost stable circular orbit ($\la$ 3$r_{\rm s}$) this ratio rises to 10 percent or higher.

\subsection{The time available for angular momentum loss}
We have shown that the radiation emitted from a black hole accretion disk can carry away a significant 
amount of angular momentum on a timescale $t_{\rm acc}$ given by equation (8). However, as shown in Section 3.1, the energy that
is radiated away is not sufficient to remove all of the angular momentum necessary to drive an accretion flow.  To understand more fully how to reconcile 
these two facts, we compare the time that a fluid element passing through the disk spends at a radius $r$ to the time $t_{\rm acc}$
that is required for its angular momentum to be carried away by radiation.  From the continuity equation, in our steady state disk model
the velocity $v_{\rm r}$ at which such a fluid element falls towards the black hole is

\begin{equation}
v_{\rm r} = \frac{\dot{M_{\rm BH}}}{2 \pi r \Sigma} \mbox{\ ,}
\end{equation}
which yields the following for the time $t_{\rm r}$ that a fluid element spends at a radius $r$ in its passage through the disk:

\begin{equation}
t_{\rm r} \sim \frac{r}{v_{\rm r}} = \frac{2 \pi r^2 \Sigma}{\dot{M_{\rm BH}}} \mbox{\ .}
\end{equation}

Taking the ratio of this time to the time $t_{\rm acc}$ in which thermal emission would carry away its angular momentum  
(equation 8) gives an estimate of the fraction of the angular momentum of the fluid element that is in fact carried away by radiation at radius $r$:

\begin{equation}
\frac{t_{\rm r}}{t_{\rm acc}} = \frac{3 G M_{\rm BH}}{4 r c^2} = \frac{3}{8}\frac{r_{\rm s}}{r} \mbox{\ .}
\end{equation}
This is in close agreement with the limit derived above from energetic considerations, and provides some 
additional \linebreak%%%%%
physical insight   
into the process at hand.  The accretion flow must be driven primarily, especially at large radii, by processes other than thermal emission, such as viscosity.  
However, as material passes through the disk a portion of the heat that it gains is radiated away, carrying with it angular momentum.
Although in general the material passes through the disk faster than its angular momentum is radiated away, as the material falls to smaller radii
it becomes hotter and more luminous, and a correspondingly larger fraction of its angular momentum is lost to radiation.

\section{Discussion and Conclusions}
We have shown that thermal emission from black hole accretion disks can remove specific angular momentum and aid in driving an accretion flow.  
In particular, we have found that the fraction of the angular momentum lost in a standard, viscous 
accretion disk that is carried away by radiation is a strong function of distance $r$ from the central black hole, namely this fraction is found to be
$\sim$ 0.4 $r_{\rm s}$/$r$.  Thus, this effect can be strong near the base of an accretion disk, 
but is of diminishing importance at larger radii.  We emphasize that, as shown in Section 3, thermal emission alone can not drive the accretion flow, and that 
the remainder of the angular momentum lost must be transported by other processes, such as turbulent viscosity. 
Furthermore, near the inner edge of the disk at $r_{\rm inner}$, 
the temperature of the accreting material will also be somewhat lower than assumed in our simple modeling, 
reduced by a factor (1-($r_{\rm inner}$/$r$)$^{\frac{1}{2}}$)$^{\frac{1}{4}}$ due to the shear
going to zero at $r_{\rm inner}$ (e.g. Pringle 1981).  This effect will act to reduce the efficacy of this process for driving the accretion flow at the smallest radii.  

We have furthermore limited ourselves to considering a disk orbiting in a Newtonian gravitational potential. While this is clearly a valid assumption 
at large radii (i.e. $r$ $>>$ $r_{\rm s}$), and our results at large radii are thus fully valid as they stand, general relativistic effects will likely impact 
our results at small radii.  While a full general relativistic calculation is beyond the scope of the present work, 
 it is nonetheless clearly of interest to understand which effects of general relativity would alter the transport of angular momentum by radiation.  
One such effect is the recapture of radiation emitted from the disk by the disk itself (e.g. Cunningham 1976); 
this can result in up to a roughly 30 percent reduction in the number of photons, and so presumably in 
the amount of angular momentum, that is radiated away from the disk.  Also, general relativistic corrections to the gravitational potential near 
the event horizon (e.g. Paczy{\' n}sky \& Wiita 1980) may enhance the efficacy of radiation for transporting angular momentum out of the disk, due to the 
circular velocity $v$ of the orbiting material increasing more quickly with decreasing radius than in the Newtonian approximation. 
Finally, while here we have focused on the role of the emitted radiation in driving the infall of gas towards the central black hole, another closely related 
general relativistic effect of the emitted radiation is to limit the spin that can be attained by the black hole itself (Thorne 1974).
Overall, while general relativistic effects which we have not treated in our simple calculation are likely to play a role at small radii (see e.g. Page \& Thorne 1974), 
our basic conclusion that radiation transports a significant amount of angular momentum out of the disk, especially at small radii, is likely to remain unchanged.  

\acknowledgements
The author is grateful to Volker Gaibler for enlightening discussions, 
to Pawan Kumar for valuable comments on an early version of this paper, and to 
William Priedhorsky for providing critical insight in the form of the energetic considerations
presented in Section 3.1.  
Finally, the author gratefully acknowledges helpful feedback from Dominik Schleicher, 
Milos \linebreak%%%%%
Milosavjlevi{\' c}, Sadegh Khochfar, Jan-Pieter Paardekooper, Charles Gammie, and Umberto Maio.

%============================================================================
%\bibliography{../BMPS/paperI.bib}
\bibliographystyle{apj}

\appendix

\begin{figure*}
\includegraphics[width=6.7in]{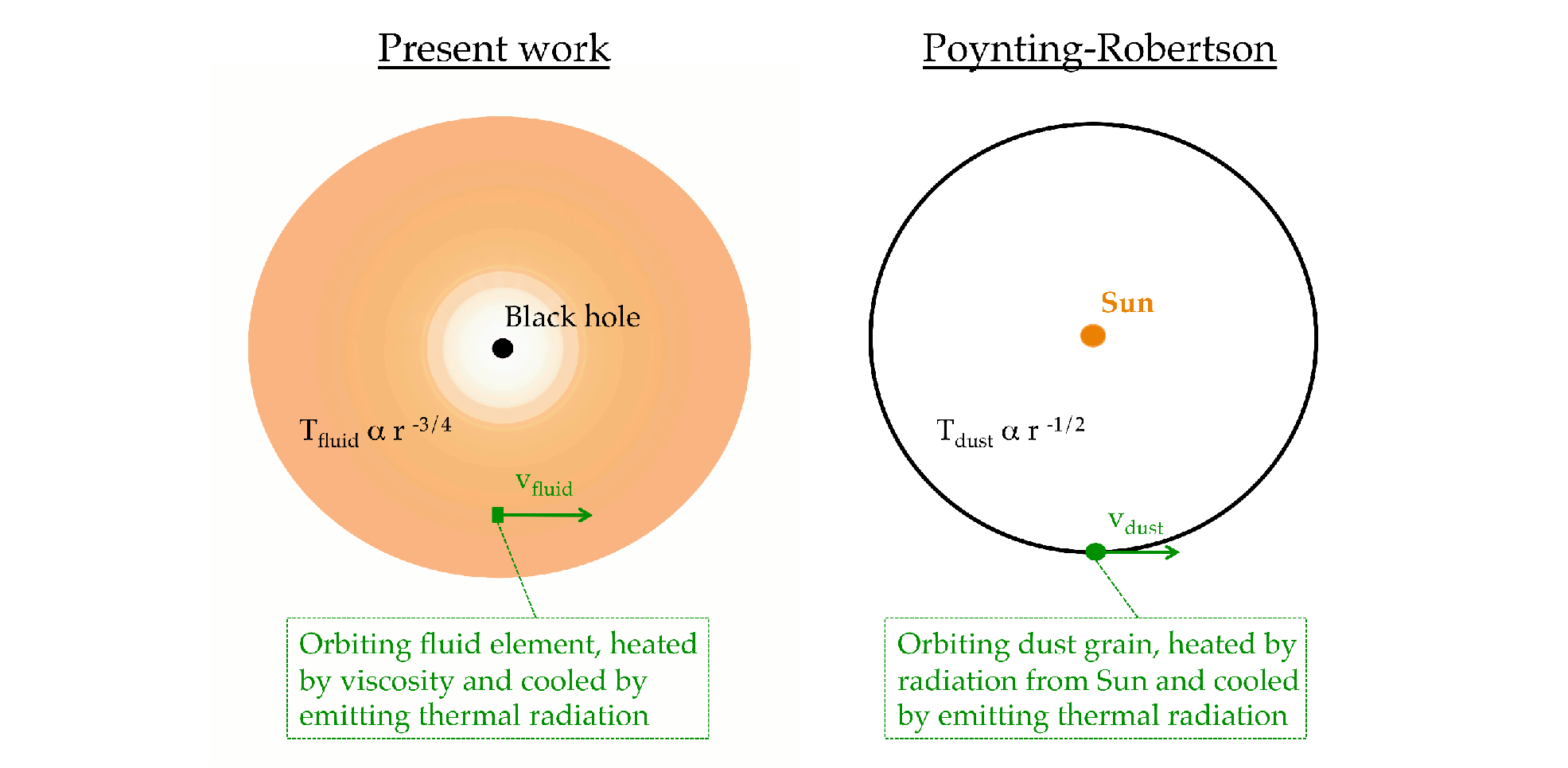}
\caption{A schematic comparison of the process considered in the present work ({\it left panel}) and the Poynting-Robertson effect ({\it right panel}).  The two are closely related in that specific angular momentum is lost via thermal radiation in each case, thereby leading to infall towards the central object, a black hole in the case of the present work and the Sun in the classical case.  The two differ in that the thermal energy that is radiated away is derived from different processes: via absorption of radiation emitted from the Sun in the classical case, and via viscous heating in the case of the standard multi-color accretion disc considered in the present work.  The infalling material in each case experiences a net heating as it falls towards the central object, with the temperature of a fluid element passing through the accretion disc going as $T_{\rm fluid}$ $\propto$ $r^{-\frac{3}{4}}$ and that of a dust grain spiraling towards the Sun going as $T_{\rm dust}$ $\propto$ $r^{-\frac{1}{2}}$.  As the source of heat is ultimately the gravitational potential energy of the infalling material, in the frame of the disk fluid element the flow of heat can not be isotropic; instead, it must be directed outward from the central black hole.  This is in essence the same situation as is realized in the Poynting-Robertson effect, and thus the outcome is the same in both cases: the anisotropically heated orbiting material loses specific angular momentum and falls inward towards the central body. 
}
\end{figure*}

\section{Comparison with the classical Poynting-Robertson effect}
It is instructive to draw a comparison between the process described in Section 2 and 
the classical Poynting-Robertson effect (Poynting 1903; Robertson 1937),
which is closely related.  In Figure A1 we make this comparison 
schematically.  Poynting and \linebreak
Robertson originally considered the process of angular momentum transport via the istropic emission of thermal
radiation by a small body (e.g. a dust grain) orbiting the Sun; in this case, the rate at which energy is radiated away is equal to the rate at 
which the grain is heated by absorption of radiation emitted by the Sun, leading to the temperature of the grain going as $T_{\rm dust}$ $\propto$ 
$r^{-\frac{1}{2}}$, assuming that it has a fixed surface area.  In the case under consideration in the present work, an isotropically emitting fluid element
in a standard multi-color black hole accretion disc is heated instead by viscosity, as described in e.g. Pringle (1981), leading to a temperature
which also increases with decreasing radius, $T_{\rm fluid}$ $\propto$ $r^{-\frac{3}{4}}$, as given by equation (1).  As such, in both cases there is a net heating of the material, 
and thus an increase of its internal energy as it falls inward (see also the general treatment of such matter-radiation interaction provided by Thomas 1930; Hsieh \& Spiegel 1976). 
This, in turn, implies that the loss of angular momentum that results from the the emission of radiation must result in a loss of specific angular momentum.   

We emphasize, as discussed explicitly in the seminal work by Robertson (1937), that 
it is the heating of the orbiting body  which results in the primary loss of specific angular momentum.  In the Poynting-Robertson case, a 
secondary effect is the aberration of the light from the central star which results in an additional, but smaller, loss of angular momentum.  While this minor effect does not operate in 
the accretion disk under consideration, the main effect is in essence the same in both cases.     

It is important, furthermore, to note that the the transport of heat in the viscous accretion disk is not isotropic in the frame of a fluid element in the disk.  If
it were, then the deposition of energy via heating would also result in the deposition of the angular momentum of the heat energy such that no net change of specific angular 
momentum would be experienced by the fluid element.  
One reason that the heating in the disk can not be isotropic is that there is a temperature gradient in the disk, and so heat 
must flow in the direction opposite this gradient; any other situation, including the isotropic transport of heat, would violate the second law of thermodynamics.   
As the temperature in the disk increases towards the central black hole, the flow of heat must be directed radially outward in the disk.  Indeed, as the ultimate source of heat in the 
disk is the gravitational potential energy of the infalling material, it  
must be the case that material is only heated by falling inwards towards the disk; in the frame of a fluid
element, this implies a flow of heat outward from the central black hole.  As this is exactly the direction in which heat flows, albeit from the central star, in the case of the 
Poynting-Robertson effect, our comparison with this effect is apt.  The result, in both cases, is a net loss of 
specific angular momentum due to heating, which in turn results in an enhanced rate of infall towards the central object.

Finally, we note that the Poynting-Robertson effect has previously been applied to accretion of gas onto compact objects as well (e.g. Blumenthal 1974; Walker \& Meszaros 1989; 
Umemura et al. 1997; Ballantyne \& Everett 2005), although for the specific case of heating by the radiation emitted by the central object.  
We emphasize that the key difference between this effect and that considered in the present work is that the source of the heat that is radiated 
away is viscous heating in the standard accretion disc model we have adopted, instead of radiation emanating from the central object.  The end result, 
however, is qualitatively the same: the radiating material in the disc loses specific angular momentum and this contributes to driving its inward spiral 
towards the black hole, as described in Section 2.

\end{document}